# Infrared nanosensors of pico- to micro-newton forces

Natalie Fardian-Melamed[1*], Artiom Skripka[2,3], Changhwan Lee[1], Benedikt Ursprung[1], Thomas P. Darlington[1], Ayelet Teitelboim[2], Xiao Qi[2], Maoji Wang[4], Jordan M. Gerton[4], Bruce E. Cohen[2,5], Emory M. Chan[2*], P. James Schuck[1*]

[1]Department of Mechanical Engineering, Columbia University, New York, New York 10027, USA
[2]The Molecular Foundry, Lawrence Berkeley National Laboratory, Berkeley, California 94720, USA
[3]Nanomaterials for Bioimaging Group, Departamento de Física de Materiales, Facultad de Ciencias, Universidad Autónoma de Madrid, Madrid 28049, Spain
[4]Department of Physics and Astronomy, University of Utah, Salt Lake City, Utah 84112, USA
[5]Division of Molecular Biophysics and Integrated Bioimaging, Lawrence Berkeley National Laboratory, Berkeley, California 94720, USA
*p.j.schuck@columbia.edu; emchan@lbl.gov; becohen@lbl.gov; natalie.melamed@columbia.edu

**Mechanical force is an essential feature for many physical and biological processes.[1-12] Remote measurement of mechanical signals with high sensitivity and spatial resolution is needed for diverse applications, including robotics,[13] biophysics,[14-20] energy storage,[21-24] and medicine.[25-27] Nanoscale luminescent force sensors excel at measuring piconewton forces,[28-32] while larger sensors have proven powerful in probing micronewton forces.[33,34] However, large gaps remain in the force magnitudes that can be probed remotely from subsurface or interfacial sites, and no individual, non-invasive sensor is capable of measuring over the large dynamic range needed to understand many systems.[35,36] Here, we demonstrate Tm$^{3+}$-doped avalanching nanoparticle[37] force sensors that can be addressed remotely by deeply penetrating near-infrared (NIR) light and can detect piconewton to micronewton forces with a dynamic range spanning more than four orders of magnitude. Using atomic force microscopy coupled with single-nanoparticle optical spectroscopy, we characterize the mechanical sensitivity of the photon avalanching process and reveal its exceptional force responsiveness. By manipulating the Tm$^{3+}$ concentrations and energy transfer within the nanosensors, we demonstrate different optical force-sensing modalities, including mechanobrightening and mechanochromism. The adaptability of these nanoscale optical force sensors, along with their multiscale sensing capability, enable operation in the dynamic and versatile environments present in real-world, complex structures spanning biological organisms to nanoelectromechanical systems (NEMS).**

Engineered and biological systems, such as integrated NEMS, developing embryos, energy storage units, and migrating cells, experience multiple scales of force as a consequence of their inherently complex, multicomponent designs that span disparate length scales.[6,15,21,38-44] The ability to study force-dependent processes in such systems is essential for understanding their central mechanisms.[20,45,46] Remotely addressable probes are needed for *in situ* detection of changes or malfunction in these processes,[23,24,27,42] particularly below surfaces or at nanoscale interfaces. At such interfaces, environmental constraints or sample fragility render direct force measurements and electrical data transmission impractical or impossible.

Optical methods excel at minimally invasive mechanical sensing.[16,35,46] Optical imaging force inference methods can deduce force-induced displacements with exquisite sensitivity, yet spatial resolution is limited by diffraction, and combination with other methods is required for retrieval of absolute force.[16] Alternatively, single luminescent nanoprobes act as reporters of their local nanoscale environments. But the dynamic ranges of forces they can sense are limited (Fig. 1a),[35,36] with no individual probe capable of accessing the multiple scales of force existing in many complex systems. Additionally, while luminescent force sensors have proven useful in understanding the molecular forces within cells, such sensors can exhibit limited photostability – restricting continuous force monitoring, or require visible wavelengths – thus inhibiting subsurface measurements.[35]

To develop a remote force sensing system that overcomes current limitations, we took advantage of unexpected observations that avalanching nanoparticles (ANPs) undergo significant changes in emission when tapped with AFM tips. ANPs[37,47-51] are a class of steeply nonlinear upconverting nanoparticles (UCNPs), which are lanthanide-based nanocrystal phosphors that convert multiple sequentially absorbed NIR photons into higher-energy emitted photons, and show no overlap with sample autofluorescence, no on-off blinking, and no measurable photobleaching, even under prolonged lasing or single-particle excitation.[52-57] Because of their steeply nonlinear relation $s$ between pump power and emission intensity, ANPs are able to amplify minute changes in signal input to giant changes in output (*i.e.*, $I_{out} = I_{in}^s$; where $I$ is intensity and nonlinear order $s > 15$).[37,47-51] Recent studies have shown that both the photon avalanche threshold and the degree of nonlinearity $s$ can be modified by manipulating lanthanide concentration.[37,47,48]

To determine the mechano-optical response of $Tm^{3+}$-doped $NaYF_4$ ANPs, we studied them as single ANPs under ambient conditions, using atomic force microscope (AFM) tips for force application, in combination with an inverted optical microscope for NIR excitation and emission measurements. To minimize quenching or energy transfer (ET) between ANP and tip or other surrounding components,[58,59] we synthesized all ANPs in this study with thick (>5 nm) undoped $NaYF_4$ shells (Supplementary Note 1, Table S1, and Fig. S2).

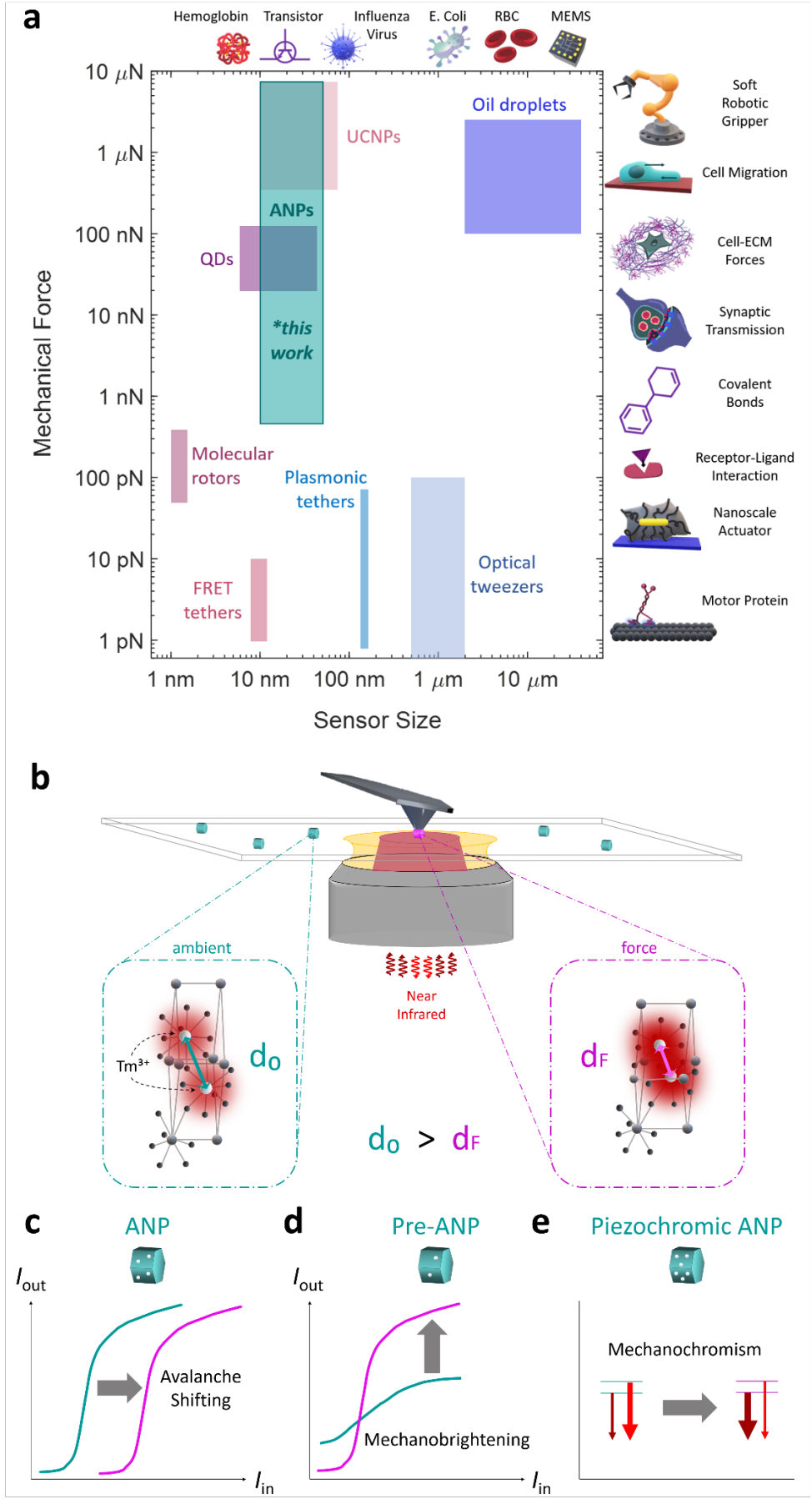

**Fig. 1: Photon avalanche-based nanoscale force sensors**. **a.** Comparison of force dynamic range and sensor size with previously reported luminescent force sensors. QD, semiconductor quantum dot; FRET, Förster resonance ET; UCNP, upconverting NP; RBC, red blood cell; ECM, extracellular matrix (sensors and force magnitude examples from ref.[2,28-36,38,60-67]). **b.** Interionic distance (*d*) decrease and consequent energy transfer (ET) and vibrational mode energy increase within a $Tm^{3+}$-doped $NaYF_4$ nanoparticle (NP) under applied force leads to three optical readout modalities (depicted in **c**, **d**, and **e**), depending on the intrinsic $Tm^{3+}$-doping concentration. **c.** Moderately doped (4.5% - 8% $Tm^{3+}$) ANPs exhibit steeply nonlinear emission *versus* excitation curves that shift with applied force, enabling observation of large emission signal change with minute applied force. **d.** Lower doped (4% $Tm^{3+}$) pre-ANPs transform from their energy-looping state to an avalanching state, prompting substantial emission enhancement with applied force. **e.** Highly doped (15% $Tm^{3+}$) piezochromic ANPs emit light from two $Tm^{3+}$ energy levels with intensities that vary differently under applied force, enabling ratiometric optical readout of applied force.

## ANP extreme nonlinearity offers access to a wide dynamic range of forces spanning 4 orders of magnitude

The lowest laser intensity at which photon avalanche is observed depends on ANP $Tm^{3+}$ concentration, shifting to higher threshold intensities as doping content increases.[37,47,48] We hypothesized that the same trend will be observed when applying force upon an ANP, as compressive force should decrease the interionic distances within the ANP, and hence effectively increase its concentration (Supplementary Note 2). In addition, a decrease in interionic distances should increase the energy of the ANP vibrational modes and hence the nonradiative rates,[68-70] further shifting the avalanche onset to higher excitation intensities[37] with application of force (as in Fig. 1c). To test this hypothesis, we measured the emission *versus* excitation for a single ANP, with and without applied force (Fig. 2a). We find that the excitation-emission curve measured at an applied force as low as 200 nN is dramatically shifted from the one measured without applying force; the same emission intensity measured at ambient force is acquired at an excitation intensity 62% larger when 200 nN are applied upon the ANP.

The drastic shift of the ANP excitation-emission curve with force implies that a large change in ANP emission will be observed per unit force for a given pump power, offering high mechano-optical sensitivity. To quantify the response, we measured the force-dependent optical emission for a series of forces, at a constant pump power. We repeated this measurement for different pump powers, in the avalanching regime (where the degree of nonlinearity *s* > 10) and in the saturation regime (where *s* ~ 2) of the single ANPs (Fig. 2b and Fig. S3-S4). We find that the mechano-optical response of an ANP is exceptionally large for all pump powers, enabling the detection of 620 pN forces in the avalanching regime and single-digit nN forces in the saturation regime (Fig. 2c-d, Supplementary Note 3, and Supplementary Methods).

To determine the dynamic range of forces measurable with an ANP, we applied 0 to 2.5 µN forces upon each ANP, and optically measured the lowest and highest detectable forces. The lowest detectable force, namely the noise-defined resolution, is shot-noise-limited and is fundamentally dependent on the intrinsic noise-equivalent sensitivity (NES)[60,71] of the ANP, and affected by integration time and setup details (Supplementary Note 3). We detect 620 pN within 3 s, and the calculated intrinsic NES of an ANP may reach 219 pN/$\sqrt{\text{Hz}}$ within the avalanching regime of the ANP (Fig. 2c, Supplementary Note 3, and Fig. S6). The highest detectable force, namely the force range, is dark-noise-limited and is fundamentally dependent on the intrinsic quantum yield of the ANP and the nanocrystal elastic limit. We detect 1.7 µN of force within the saturation regime of these ANP compositions. We observe no optical degradation resulting from plastic deformation or photobleaching, for up to 25 compression cycles measured within this 2.5 µN force range and these measurement times and pump powers (Supplementary Note 4 and Fig. S5). We find that the highly nonlinear mechano-optical response, which is maximized at the avalanching regime, enables the detection of hundreds of pN to hundreds of nN forces at lower pump powers. Meanwhile, the ANP high emission, which is maximized at the saturation regime, allows the detection of single-digit nN to single-digit µN at higher pump intensities (Fig. 2c-d). The same large dynamic range spanning three orders of magnitude of force is maintained throughout all pump intensities for all single ANPs measured (Fig. 2c).

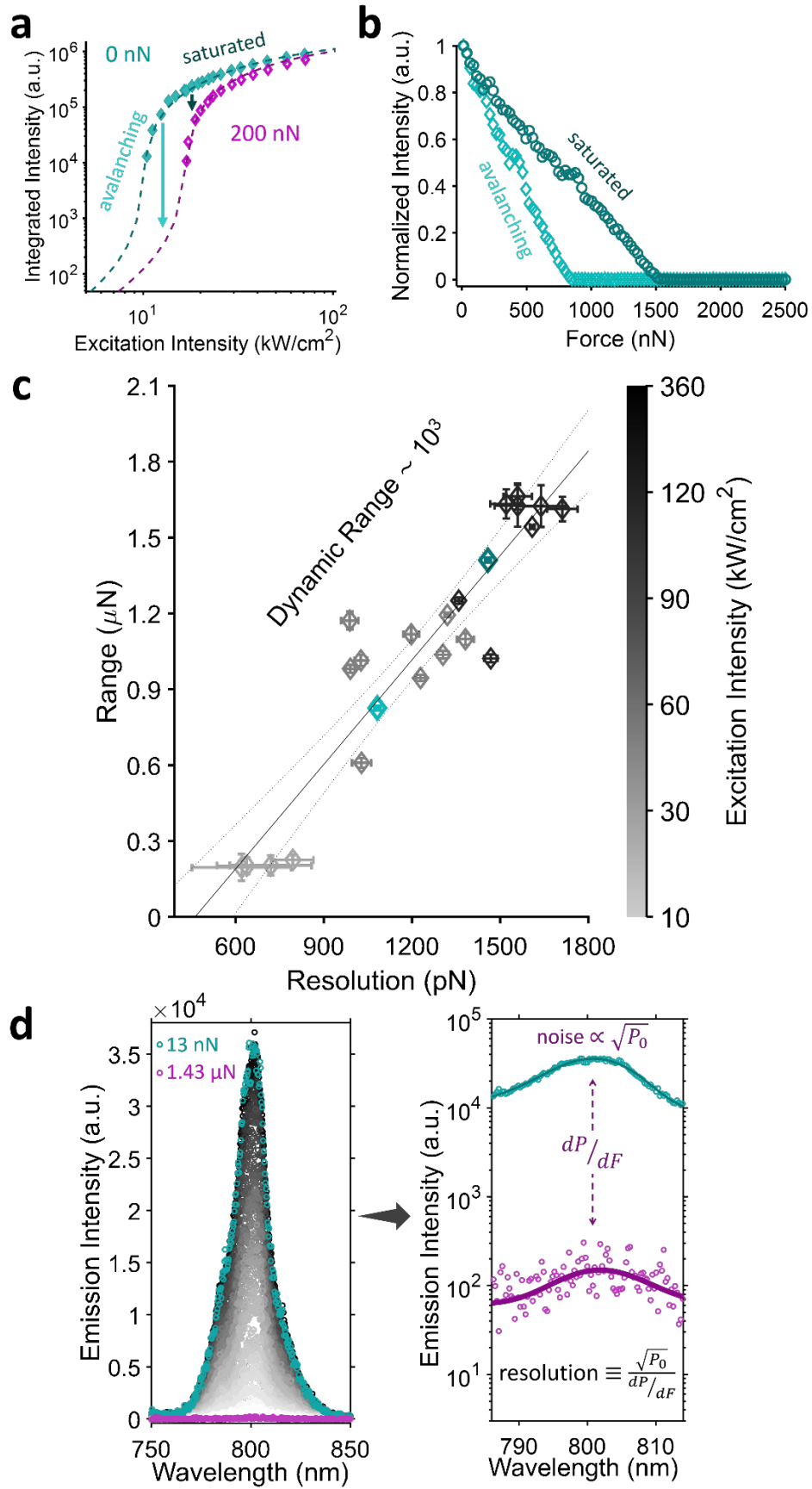

**Fig. 2: Mechano-optical characterization of single ANPs. a.** Emission (at 800 nm) *versus* excitation (at 1064 nm) intensities measured for a single ANP with and without applied force. The dashed lines are photon avalanche differential rate equation model[37] fits (Table S2). The larger (smaller) change in signal per unit force, represented by the left (right) arrow, occurs in the avalanching (saturation) regime of the ANP. **b.** Emission intensity as a function of applied force for a single ANP excited in its avalanching regime and saturation regime. **c.** Highest detectable force (range) and lowest detectable force (resolution) for 23 single-ANP compression cycles for seven ANPs at different excitation intensities. Standard errors are derived from the 95% confidence bounds on the linear fits to the emission intensity *versus* applied force graphs (as in **b**), each containing 10-200 data points. The two compression cycles from the single ANP in **b** are enlarged and color-coded as in **b** herein. **d.** Emission spectra (*left*) and their peak magnifications (*right*) measured at the start and end of a force-emission compression cycle, for an ANP excited in its saturation regime. *Left:* 30 force-dependent spectra sequentially measured in between the start (teal) and end (magenta) of the compression cycle are depicted in grey. *Right:* The noise-defined resolution (*P* – photon counts, *F* – force), along with fits to sums of Gaussians centered at the corresponding $Tm^{3+}$ transitions within the wavelength range, are depicted. Integration time is 3 s for all measurements. $Tm^{3+}$ concentrations are 4.5% (**a**) and 7% (**b**-**d**).

**Mechanobrightening through UCNP-to-ANP transformation**

Motivated by these results, we hypothesized that photon avalanching behavior could be initiated by applied force, resulting in mechanobrightening nanosensors. We envisioned creating non-avalanching UCNPs with $Tm^{3+}$ concentrations just below those sufficient to attain photon-avalanche at ambient conditions, and transforming them into avalanching UCNPs through the utilization of force.

It was previously found that ANPs with 8% $Tm^{3+}$ show a signature photon avalanching response.[37] But for lower concentrations of 1% to 4%, a threshold intensity was less apparent, and the nonlinearities observed in the emission *versus* excitation curves were much smaller.[37] We posited that if the effective $Tm^{3+}$ concentration of a 'pre-avalanching nanoparticle' – or 'pre-ANP' for short – were increased *via* applied force, one would render the NP avalanching under pressure. Since photon-avalanche is sustained through efficient cross-relaxation (a form of ET) between the emitting lanthanides,[37] which is also influenced by phonon-mediated processes, application of force should enhance this process, and hence transform a pre-ANP into an ANP. We therefore designed a set of particles with $Tm^{3+}$ concentrations ranging from 7% down to 4%, in search of the pre-avalanching concentration. We find that when pressing upon particles with $Tm^{3+}$ concentrations down to 4.5%, the emission decreases with applied force. In addition, 4.5% $Tm^{3+}$ NPs exhibit steeply nonlinear emission *versus* excitation (Fig. 2a and Table S2), deeming them ANPs at ambient conditions. However, when pressing upon 4% $Tm^{3+}$ particles, which do not avalanche at ambient conditions (Fig. S7), the emission intensity *increases* and is enhanced four-fold for a single force ramp of only ~400 nN (Fig. 3a).

To test the viability of mechanobrightening for force sensing, we repeatedly applied forces of 0 to 2.5 µN upon single pre-ANPs, excited at various pump powers (Fig. S8-S10). We find that the transformation into an ANP occurs at ~400 nN, after which the now-ANP follows conventional ANP response to force – that of emission decrease. This yields a mechanobrightening force range of ~400 nN for pre-ANPs. The steep increase in emission per unit force enables the detection of forces as low as 475 pN within 3 s (Fig. 3a, inset), yielding an intrinsic noise-equivalent sensitivity of 168 pN/$\sqrt{\text{Hz}}$ for pre-ANPs (Supplementary Note 3). The ability to detect forces at such high sensitivities, over wide ranges, with an amplifying emission signal, along with the robustness and repeatability of the signal under continuous pumping and force application (Fig. 3b) – make pre-ANPs highly attractive for remote mechanical sensing.

To further understand how force transforms pre-ANPs into ANPs, we characterized the emission *versus* excitation profile of a pre-ANP with and without applied force. Unlike ANP excitation-emission curves, which shift to the right upon application of force (Fig. 2a and Fig. 1c) – pre-ANP excitation-emission curves shift upwards, with increased nonlinearity, manifested in the larger excitation-emission curve slope (Fig. 3c and Fig. 1d). This distinct change in the excitation-emission profile, observed only for pre-ANPs under applied force, highlights their transition from pre-avalanching (Fig. S7) to avalanching.

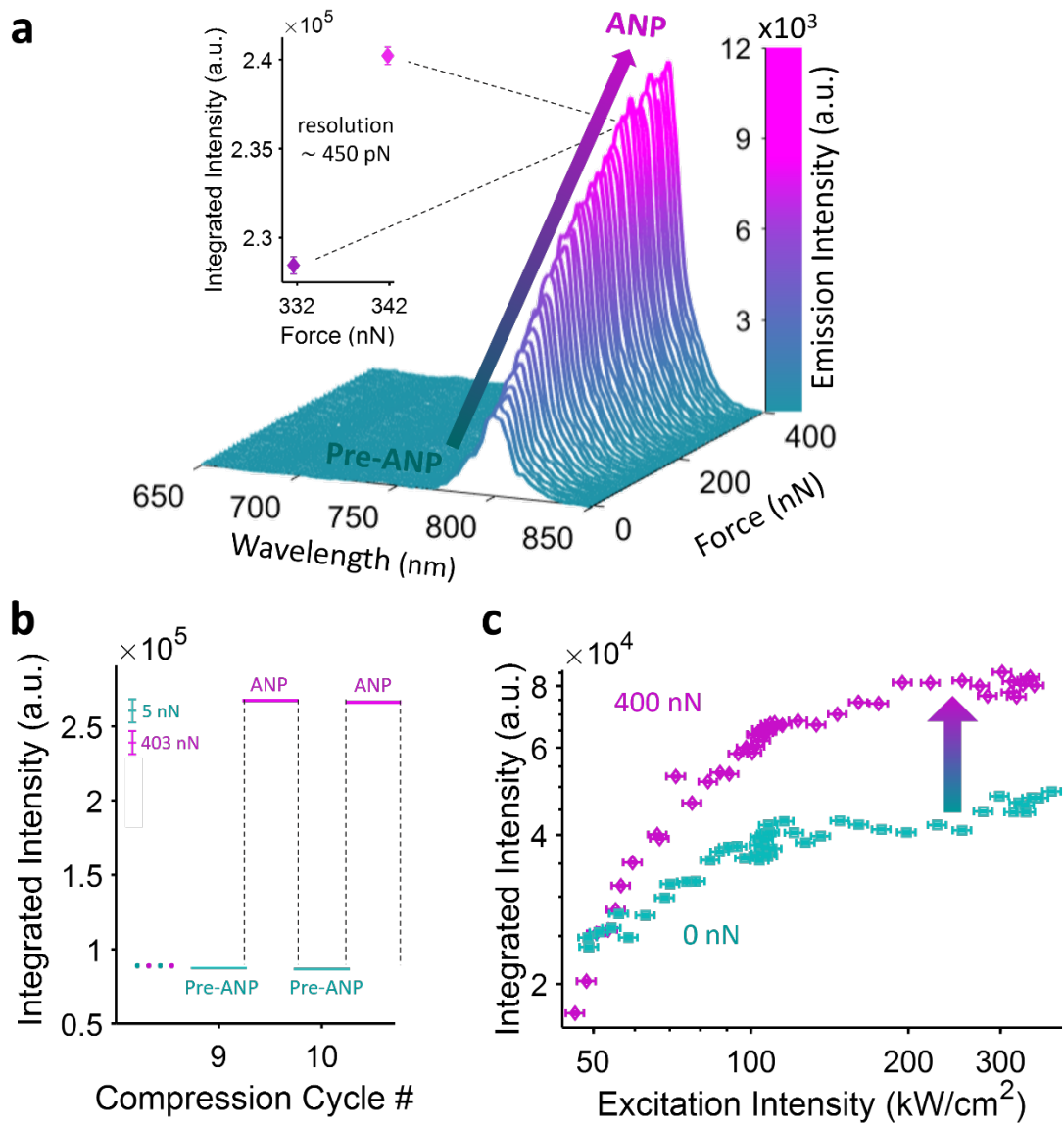

**Figure 3: Mechano-optical characterization of single pre-ANPs. a.** Emission spectra as a function of applied force for a single 4% Tm$^{3+}$ pre-ANP. The NP transforms from a pre-ANP at ambient force, to an ANP under 400 nN applied force. *Inset:* Integrated intensities from two sequentially measured emission spectra collected during an emission-*versus*-applied force scan. The noise-defined resolution (as in Fig. 2d) is depicted. **b.** Emission intensity as a function of force for a single pre-ANP after undergoing 8 compression cycles, demonstrating reproducibility. Note that the line point thicknesses are larger than the error bars for these measurements. **c.** Emission (at 800 nm) *versus* excitation (at 1064 nm) intensities measured for a single pre-ANP with and without applied force. All error bars represent the (Poisson distribution-) standard deviation of the measured signals. Each full compression cycle consists of 75 force-emission measurements, and lasts 4 min. Integration time is 3 s for all measurements.

**Mechanochromic nanosensors**

With a goal of expanding utility for a broader range of applications, we next set out to design mechanochromic nanosensors, with a dual-wavelength ratiometric readout of force. The aim was a sensor for which the signal does not depend solely on the intensity of one emission wavelength, but of two emission wavelengths that each show a different force response. The wavelengths should be spectrally proximal to enable their simultaneous detection, yet far enough apart so as not to overlap, yielding force-dependent signal that depends on the ratio of these two emissions (the 'color' of emission). Mechanochromic self-calibrated signals of this type provide a built-in control against environmental interference, which can easily disrupt single-wavelength intensity readout.[72]

To design mechanochromic nanosensors with substantial emission from both the main photon-avalanche level ($^3H_4$) and a nearby energy level ($^3F_3$), we increased the $Tm^{3+}$ concentration within the ANPs. We find that for 15% $Tm^{3+}$, there is sizeable emission at 700 nm ($^3F_3$ level) and 800 nm ($^3H_4$ level), (Fig. 4a and Fig. S11) and hereafter refer to 15% $Tm^{3+}$ NPs as 'piezochromic ANPs'.

When subjected to force, the relative emission from the $^3H_4$ level increases compared to that from the $^3F_3$ level (Fig. 4a), and the overall emission from a piezochromic ANP decreases with force (Fig. S12) as with the lower-$Tm^{3+}$-content ANPs (Fig. 2). The ratio between these two emission lines, therefore, reports on the applied force. The mechano-optical response, manifested in the percent change of the 800-nm-emission/700-nm-emission ratio per unit force, is identical for different single NPs probed during different integration times (Fig. 4b and Fig. S13), and is on average 54 %/µN (Fig. 4c). For each piezochromic ANP, the response, ambient-force brightness, and ambient-force dual-emission ratio remain unchanged under continuous pumping at high excitation intensities and after repeated compression to 2.5 µN (Fig. S14-S16). The large mechano-optical response and ambient-force brightness, together with the low ambient-force dual-emission ratio found for piezochromic ANPs (Fig. S15-S17) enable the detection of single-digit nN forces within less than 1 min (Fig. 4d), and yield an intrinsic noise-equivalent sensitivity of 3.06 nN/$\sqrt{Hz}$ (Supplementary Note 3 and Fig. S17).

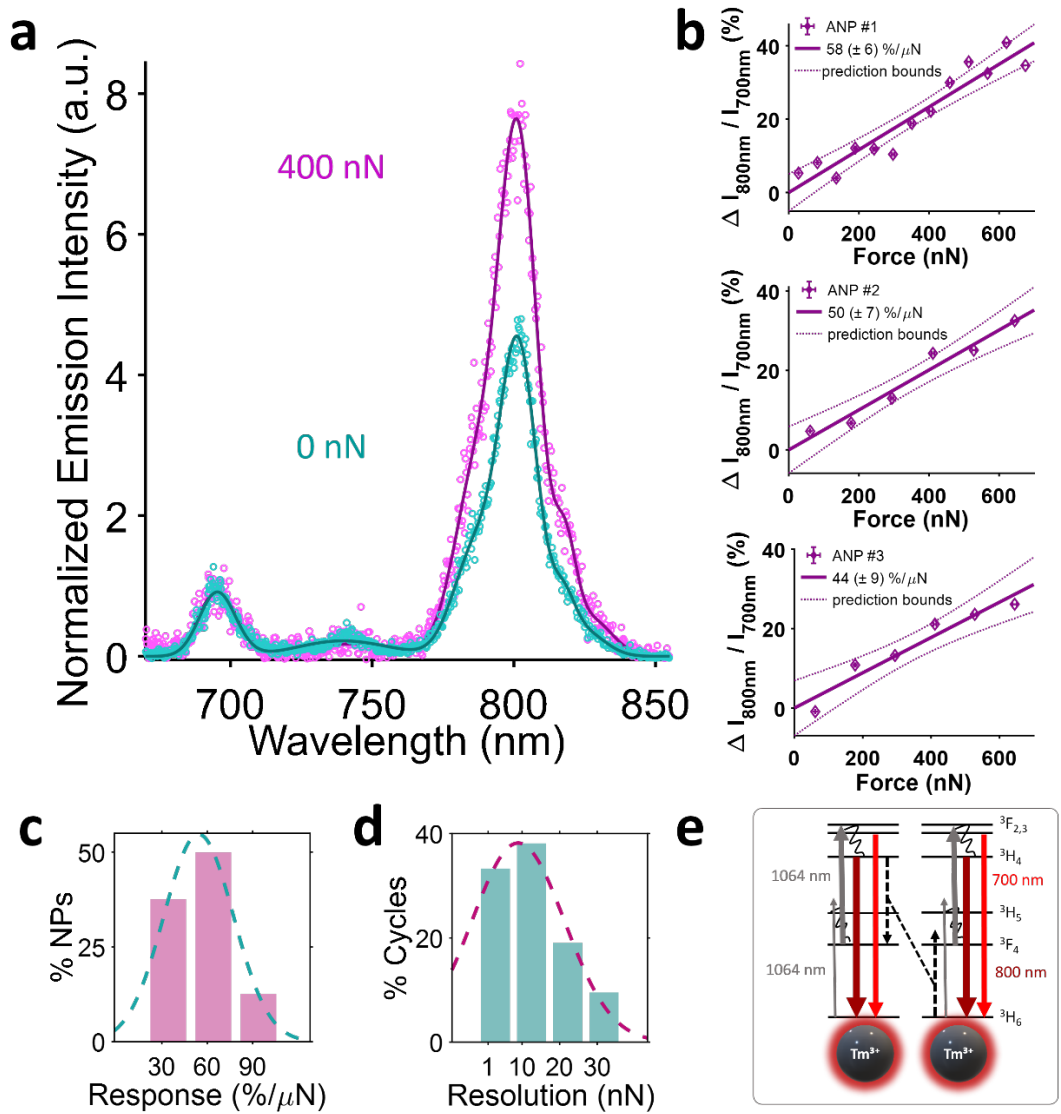

**Figure 4: Mechano-optical characterization of single piezochromic ANPs. a.** Emission spectra collected from a single 15% $Tm^{3+}$ piezochromic ANP with and without applied force, normalized to the $^3F_3$ (700 nm) emission intensity. Spectra are fit to sums of Gaussians centered at the respective $Tm^{3+}$ transitions within the spectral range. **b.** Percent change of the 800nm-emission/700nm-emission ratio *versus* applied force, for three compression cycles originating from three different piezochromic ANPs. Integration times are 50 s (ANP #1), 20 s (ANP #2), and 30 s (ANP #3). **c.** Mechano-optical response distribution (response = 54 ± 7 %/μN; data mean ± S.E.M.). **d.** Force resolution distribution (resolution = 13 ± 2 nN; data mean ± S.E.M.). Data in **c** and **d** are derived from 21 compression cycles (each consisting of 10-1000 emission *versus* force data points) on eight piezochromic ANPs. Integration times are 3 - 50 s. Dashed lines are Gaussian fits to the histograms. **e.** Main electronic levels and transitions involved when considering piezochromic dual-wavelength emission. Wavy and dashed lines represent multi-phonon relaxation and cross-relaxation, respectively.

**Discussion**

The steep nonlinearity of the photon-avalanche process allows for the amplification of subtle interionic distance changes through numerous cycles of cross-relaxation, each of which scales inversely with distance to the sixth power and/or exponentially if Dexter ET is present. This, in addition to the high intrinsic quantum yield of ANPs,[37] enables the detection of three orders of magnitude of force at any pump intensity utilized. Overall, we demonstrate that four or more orders of magnitude of force can be probed with the same single ANP by simply adjusting the pump intensity – allowing for remote multiple-scale force sensing of nanoscale environments.

The ability to transform a conventional UCNP into an ANP and back through the application and release of force, respectively, further highlights the strong dependence of the photon-avalanche mechanism on interionic distance and phonon energy. Only 400 nN of force are required to observe four-fold emission increase at ambient conditions. Although emission intensity increase with pressure has been shown in previous hydrostatic pressure studies of lanthanide-doped UCNPs,[68,73] an enhancement of 1.7x was observed at most, using orders of magnitude larger forces. Those changes in signal were attributed to modifications of the crystal field symmetry, as the NPs under study were of cubic-phase, and hence centro-symmetric in nature. Compression of hexagonal-phase NPs in those studies did not yield measurable increase in signal with pressure. Because the pre-ANPs in this study are hexagonal-phase NPs, we speculate that the observed changes rely more on interionic distance than on crystal field symmetry. As the interionic distance and phonon energy govern the degree of cross-relaxation between emitters – the main factor in photon avalanching – minute changes in the former will lead to substantial changes in signal intensity. These observations have led us to design mechanobrightening pre-ANPs, which are capable of detecting a wide range of forces, from 475 pN to 400 nN.

The piezochromic ANPs designed here offer force sensitivities an order of magnitude higher than the state of the art nanoscale ratiometric force sensors at ambient conditions.[60,71] Yet pN-scale forces, easily detectable with pre-ANPs and ANPs, were not detected with piezochromic ANPs with the same experimental conditions (Fig. S18). We hypothesize that this is due to a different underlying mechanism behind the mechanochromism ratiometric change: While the record sensitivities of pre-ANPs and ANPs are highly dependent on the emitter-emitter interionic distance and photon-avalanche magnification, the sensitivity of piezochromic ANPs is mainly dependent on *intra*-ion energy relaxation, which depends more on the overall interionic distance of the host lattice ions.[70] Once the latter is decreased by application of force, the vibrational energy of the host lattice, and therefore the nonradiative relaxation rate between the $^3F_3$ level and the $^3H_4$ level, is increased, as seen in previous studies of lanthanide-doped $NaYF_4$.[68,69,74]

To conclude, by leveraging the high nonlinearity offered by photon-avalanche, we design nanoscale, remotely controlled, NIR-input, NIR-output, high dynamic range, force sensors with force sensitivities falling within the previously inaccessible force range of all existing optical nanosensors.[35,60] Different modalities of readout (from mechanobrightening to mechanochromism), with different force resolutions (from pNs to nNs), can be utilized by choice of nanosensor $Tm^{3+}$ concentration. The ability to remotely and accurately quantify local forces on these multiple scales will enable advances in both fundamental investigations and critical applications, allowing discovery and precision study of local mechano-induced processes and, ultimately, their quantification, nano-mapping of spatial distributions, and early detection of malfunction, in technological devices and physiology.

**References**

1   Gouveia, B. *et al.* Capillary forces generated by biomolecular condensates. *Nature* **609**, 255-264 (2022).
2   Ucar, H. *et al.* Mechanical actions of dendritic-spine enlargement on presynaptic exocytosis. *Nature* **600**, 686-689 (2021).
3   Handler, A. & Ginty, D. D. The mechanosensory neurons of touch and their mechanisms of activation. *Nature Reviews Neuroscience* **22**, 521-537 (2021).
4   Qiu, X. & Müller, U. Sensing sound: Cellular specializations and molecular force sensors. *Neuron* (2022).
5   Marshall, K. & Patapoutian, A. Getting a grip on touch receptors. *Science* **368**, 1311-1312 (2020).
6   Vining, K. H. & Mooney, D. J. Mechanical forces direct stem cell behaviour in development and regeneration. *Nature reviews Molecular cell biology* **18**, 728-742 (2017).
7   Nonomura, K. *et al.* Piezo2 senses airway stretch and mediates lung inflation-induced apnoea. *Nature* **541**, 176-181 (2017).
8   Murthy, S. E., Dubin, A. E. & Patapoutian, A. Piezos thrive under pressure: mechanically activated ion channels in health and disease. *Nature reviews Molecular cell biology* **18**, 771-783 (2017).
9   Lam, R. M. *et al.* PIEZO2 and perineal mechanosensation are essential for sexual function. *Science* **381**, 906-910 (2023).
10  Hill, R. Z., Loud, M. C., Dubin, A. E., Peet, B. & Patapoutian, A. PIEZO1 transduces mechanical itch in mice. *Nature* **607**, 104-110 (2022).
11  Marshall, K. L. *et al.* PIEZO2 in sensory neurons and urothelial cells coordinates urination. *Nature* **588**, 290-295 (2020).
12  Zeng, W.-Z. *et al.* PIEZOs mediate neuronal sensing of blood pressure and the baroreceptor reflex. *Science* **362**, 464-467 (2018).
13  Li, M., Pal, A., Aghakhani, A., Pena-Francesch, A. & Sitti, M. Soft actuators for real-world applications. *Nature Reviews Materials* **7**, 235-249 (2022).

14	Romani, P., Valcarcel-Jimenez, L., Frezza, C. & Dupont, S. Crosstalk between mechanotransduction and metabolism. *Nature Reviews Molecular Cell Biology* **22**, 22-38 (2021).
15	Saraswathibhatla, A., Indana, D. & Chaudhuri, O. Cell–extracellular matrix mechanotransduction in 3D. *Nature Reviews Molecular Cell Biology*, 1-22 (2023).
16	Gómez-González, M., Latorre, E., Arroyo, M. & Trepat, X. Measuring mechanical stress in living tissues. *Nature Reviews Physics* **2**, 300-317 (2020).
17	Jain, S. *et al.* The role of single-cell mechanical behaviour and polarity in driving collective cell migration. *Nature physics* **16**, 802-809 (2020).
18	De Belly, H., Paluch, E. K. & Chalut, K. J. Interplay between mechanics and signalling in regulating cell fate. *Nature Reviews Molecular Cell Biology* **23**, 465-480 (2022).
19	Qiu, Y., Myers, D. R. & Lam, W. A. The biophysics and mechanics of blood from a materials perspective. *Nature Reviews Materials* **4**, 294-311 (2019).
20	Killeen, A., Bertrand, T. & Lee, C. F. Polar fluctuations lead to extensile nematic behavior in confluent tissues. *Physical Review Letters* **128**, 078001 (2022).
21	de Vasconcelos, L. S. *et al.* Chemomechanics of Rechargeable Batteries: Status, Theories, and Perspectives. *Chemical Reviews* **122**, 13043-13107, doi:10.1021/acs.chemrev.2c00002 (2022).
22	Huang, W. *et al.* Onboard early detection and mitigation of lithium plating in fast-charging batteries. *Nature Communications* **13**, 7091, doi:10.1038/s41467-022-33486-4 (2022).
23	Liu, K., Liu, Y., Lin, D., Pei, A. & Cui, Y. Materials for lithium-ion battery safety. *Science Advances* **4**, eaas9820, doi:doi:10.1126/sciadv.aas9820 (2018).
24	Doux, J. M. *et al.* Stack pressure considerations for room-temperature all-solid-state lithium metal batteries. *Advanced Energy Materials* **10**, 1903253 (2020).
25	Kaushik, S. & Persson, A. I. Unlocking the dangers of a stiffening brain. *Neuron* **100**, 763-765 (2018).
26	Chen, X. *et al.* A feedforward mechanism mediated by mechanosensitive ion channel PIEZO1 and tissue mechanics promotes glioma aggression. *Neuron* **100**, 799-815. e797 (2018).
27	Zhang, J. & Reinhart-King, C. A. Targeting tissue stiffness in metastasis: mechanomedicine improves cancer therapy. *Cancer Cell* **37**, 754-755 (2020).
28	Nickels, P. C. *et al.* Molecular force spectroscopy with a DNA origami–based nanoscopic force clamp. *Science* **354**, 305-307 (2016).
29	Ringer, P. *et al.* Multiplexing molecular tension sensors reveals piconewton force gradient across talin-1. *Nature methods* **14**, 1090-1096 (2017).
30	Stabley, D. R., Jurchenko, C., Marshall, S. S. & Salaita, K. S. Visualizing mechanical tension across membrane receptors with a fluorescent sensor. *Nature methods* **9**, 64-67 (2012).
31	Brockman, J. M. *et al.* Live-cell super-resolved PAINT imaging of piconewton cellular traction forces. *Nature methods* **17**, 1018-1024 (2020).
32	Blanchard, A. T. & Salaita, K. Emerging uses of DNA mechanical devices. *Science* **365**, 1080-1081 (2019).
33	Campàs, O. *et al.* Quantifying cell-generated mechanical forces within living embryonic tissues. *Nature methods* **11**, 183-189 (2014).
34	Serwane, F. *et al.* In vivo quantification of spatially varying mechanical properties in developing tissues. *Nature methods* **14**, 181-186 (2017).
35	Mehlenbacher, R. D., Kolbl, R., Lay, A. & Dionne, J. A. Nanomaterials for in vivo imaging of mechanical forces and electrical fields. *Nature Reviews Materials* **3**, 1-17 (2017).
36	Sun, W., Gao, X., Lei, H., Wang, W. & Cao, Y. Biophysical approaches for applying and measuring biological forces. *Advanced Science* **9**, 2105254 (2022).
37	Lee, C. *et al.* Giant nonlinear optical responses from photon-avalanching nanoparticles. *Nature* **589**, 230-235 (2021).

38    Boocock, D., Hino, N., Ruzickova, N., Hirashima, T. & Hannezo, E. Theory of mechanochemical patterning and optimal migration in cell monolayers. *Nature physics* **17**, 267-274 (2021).
39    Miroshnikova, Y. A. *et al.* Adhesion forces and cortical tension couple cell proliferation and differentiation to drive epidermal stratification. *Nature cell biology* **20**, 69-80 (2018).
40    Petridou, N. I., Spiró, Z. & Heisenberg, C.-P. Multiscale force sensing in development. *Nature cell biology* **19**, 581-588 (2017).
41    Van Helvert, S., Storm, C. & Friedl, P. Mechanoreciprocity in cell migration. *Nature cell biology* **20**, 8-20 (2018).
42    Liu, C. *et al.* Heterogeneous microenvironmental stiffness regulates pro-metastatic functions of breast cancer cells. *Acta Biomaterialia* **131**, 326-340 (2021).
43    Midolo, L., Schliesser, A. & Fiore, A. Nano-opto-electro-mechanical systems. *Nature nanotechnology* **13**, 11-18 (2018).
44    Tsoukalas, K., Lahijani, B. V. & Stobbe, S. Impact of transduction scaling laws on nanoelectromechanical systems. *Physical Review Letters* **124**, 223902 (2020).
45    Polacheck, W. J. & Chen, C. S. Measuring cell-generated forces: a guide to the available tools. *Nature methods* **13**, 415-423 (2016).
46    Wu, J., Lewis, A. H. & Grandl, J. Touch, tension, and transduction–the function and regulation of Piezo ion channels. *Trends in biochemical sciences* **42**, 57-71 (2017).
47    Bednarkiewicz, A., Chan, E. M., Kotulska, A., Marciniak, L. & Prorok, K. Photon avalanche in lanthanide doped nanoparticles for biomedical applications: super-resolution imaging. *Nanoscale Horizons* **4**, 881-889 (2019).
48    Dudek, M. *et al.* Size-Dependent Photon Avalanching in Tm3+ Doped LiYF4 Nano, Micro, and Bulk Crystals. *Advanced Optical Materials* **10**, 2201052 (2022).
49    Skripka, A. *et al.* A Generalized Approach to Photon Avalanche Upconversion in Luminescent Nanocrystals. *Nano Letters* **23**, 7100-7106, doi:10.1021/acs.nanolett.3c01955 (2023).
50    Liang, Y. *et al.* Migrating photon avalanche in different emitters at the nanoscale enables 46th-order optical nonlinearity. *Nature Nanotechnology* **17**, 524-530 (2022).
51    Zhang, Z. *et al.* Tuning phonon energies in lanthanide-doped potassium lead halide nanocrystals for enhanced nonlinearity and upconversion. *Angewandte Chemie International Edition* **62**, e202212549 (2023).
52    Wu, S. *et al.* Non-blinking and photostable upconverted luminescence from single lanthanide-doped nanocrystals. *Proceedings of the National Academy of Sciences* **106**, 10917-10921 (2009).
53    Park, Y. I. *et al.* Nonblinking and nonbleaching upconverting nanoparticles as an optical imaging nanoprobe and T1 magnetic resonance imaging contrast agent. *Advanced Materials* **21**, 4467-4471 (2009).
54    Cohen, B. E. Beyond fluorescence. *Nature* **467**, 407-408 (2010).
55    Ostrowski, A. D. *et al.* Controlled synthesis and single-particle imaging of bright, sub-10 nm lanthanide-doped upconverting nanocrystals. *ACS nano* **6**, 2686-2692 (2012).
56    Gargas, D. J. *et al.* Engineering bright sub-10-nm upconverting nanocrystals for single-molecule imaging. *Nature nanotechnology* **9**, 300-305 (2014).
57    Lee, C. *et al.* Indefinite and bidirectional near-infrared nanocrystal photoswitching. *Nature* **618**, 951-958 (2023).
58    Fischer, S., Bronstein, N. D., Swabeck, J. K., Chan, E. M. & Alivisatos, A. P. Precise tuning of surface quenching for luminescence enhancement in core–shell lanthanide-doped nanocrystals. *Nano letters* **16**, 7241-7247 (2016).
59    Johnson, N. J. *et al.* Direct evidence for coupled surface and concentration quenching dynamics in lanthanide-doped nanocrystals. *Journal of the American Chemical Society* **139**, 3275-3282 (2017).

60  Casar, J. R., McLellan, C. A., Siefe, C. & Dionne, J. A. Lanthanide-based nanosensors: refining nanoparticle responsiveness for single particle imaging of stimuli. *ACS photonics* **8**, 3-17 (2020).
61  Sinatra, N. R. *et al.* Ultragentle manipulation of delicate structures using a soft robotic gripper. *Science Robotics* **4**, eaax5425 (2019).
62  Trepat, X. *et al.* Physical forces during collective cell migration. *Nature physics* **5**, 426-430 (2009).
63  Huber, M., Casares-Arias, J., Fässler, R., Müller, D. J. & Strohmeyer, N. In mitosis integrins reduce adhesion to extracellular matrix and strengthen adhesion to adjacent cells. *Nature Communications* **14**, 2143 (2023).
64  Fletcher, D. A. & Mullins, R. D. Cell mechanics and the cytoskeleton. *Nature* **463**, 485-492 (2010).
65  Feld, L. *et al.* Cellular contractile forces are nonmechanosensitive. *Science advances* **6**, eaaz6997 (2020).
66  Liu, Z. *et al.* Nanoscale optomechanical actuators for controlling mechanotransduction in living cells. *Nature methods* **13**, 143-146 (2016).
67  Van den Heuvel, M. G. & Dekker, C. Motor proteins at work for nanotechnology. *Science* **317**, 333-336 (2007).
68  Wisser, M. D. *et al.* Strain-induced modification of optical selection rules in lanthanide-based upconverting nanoparticles. *Nano letters* **15**, 1891-1897 (2015).
69  Lage, M. M., Moreira, R. L., Matinaga, F. M. & Gesland, J.-Y. Raman and infrared reflectivity determination of phonon modes and crystal structure of Czochralski-grown $NaLnF_4$ (Ln= La, Ce, Pr, Sm, Eu, and Gd) single crystals. *Chemistry of materials* **17**, 4523-4529 (2005).
70  van Swieten, T. P. *et al.* Extending the dynamic temperature range of Boltzmann thermometers. *Light: Science & Applications* **11**, 1-11 (2022).
71  McLellan, C. A. *et al.* Engineering Bright and Mechanosensitive Alkaline-Earth Rare-Earth Upconverting Nanoparticles. *The Journal of Physical Chemistry Letters* **13**, 1547-1553 (2022).
72  Park, S.-H., Kwon, N., Lee, J.-H., Yoon, J. & Shin, I. Synthetic ratiometric fluorescent probes for detection of ions. *Chemical Society Reviews* **49**, 143-179 (2020).
73  Runowski, M. *et al.* Lifetime nanomanometry–high-pressure luminescence of up-converting lanthanide nanocrystals–$SrF_2$: $Yb^{3+}$, $Er^{3+}$. *Nanoscale* **9**, 16030-16037 (2017).
74  Dong, H., Sun, L.-D. & Yan, C.-H. Local structure engineering in lanthanide-doped nanocrystals for tunable upconversion emissions. *Journal of the American Chemical Society* **143**, 20546-20561 (2021).

**Data availability**

All data generated or analyzed during this study, which support the plots within this paper and other findings of this study, are included in this published article and its Supplementary Information. Further data are available from the corresponding authors upon reasonable request.

**Acknowledgements:** The authors thank Kevin W. C. Kwock and Robert G. Stark for assistance with the setup configuration. N.F.-M. gratefully acknowledges support from the European Union's Horizon 2020 research and innovation program under the Marie Skłodowska-Curie grant agreement No. 893439, the US Department of State Fulbright Scholarship Program, the Zuckerman-CHE STEM Leadership Program, the Israel Scholarship Education Foundation (ISEF) International Fellowship Program, and the Weizmann Institute's Women's Postdoctoral Career Development Award. B.U. and P.J.S. acknowledge support by the National Science Foundation under grant no. CHE-2203510. A.S. acknowledges the support from the European Union's Horizon 2020 research and innovation program under the Marie Skłodowska-Curie grant agreement No. 895809 (MONOCLE). Work at the Molecular Foundry was supported by the Office of Science, Office of Basic Energy Sciences, of the US Department of Energy under contract number DE-AC02-05CH11231. X.Q., B.E.C., and E.M.C. were supported in part by the Defense Advanced Research Projects Agency (DARPA) ENVision program under contract HR0011257070, and C.L. and P.J.S. under DARPA ENVision contract HR00112220006. T.P.D. and P.J.S. also acknowledge support from Programmable Quantum Materials, an Energy Frontier Research Center funded by the US DOE, Office of Science, Basic Energy Sciences (BES), under award DE-SC0019443.

**Author Contributions:** Conceptualization: N.F.-M. and P.J.S. Experimental design: N.F.-M. and P.J.S. Experimental setup: N.F.-M., B.U., T.P.D., and C.L. Mechano-optical measurements: N.F.-M., M.W., J.M.G., and P.J.S. Mechano-optical data analysis: N.F.-M., B.U., and C.L. Advanced nanocrystal synthesis and structural characterization: A.S., A.T., X.Q., B.E.C., and E.M.C. Single-nanocrystal sample preparation: N.F.-M. and C.L. Manuscript writing: All Authors.

**Competing Interests:** The authors declare no competing interests.

**Supplementary Information:**

Materials and Methods

Supplementary Notes 1 – 4

Supplementary Tables S1 – S2

Supplementary Figures S1 – S18

Supplementary References